# Asteroids in the service of humanity

## I. A. Crawford

Department of Earth and Planetary Sciences
Birkbeck College, University of London

There are at least three compelling reasons for the human race to initiate a major programme to explore and better understand the 'minor planets' of the Solar System:

(1) Enhancing scientific knowledge: Asteroids represent material that failed to get incorporated into planets when the Solar System formed. As such they constitute a metaphorical 'gold mine' of scientific information relating to the properties of the early Sun, astrophysical processes in the protoplanetary disk, and the early stages of planetesimal formation and evolution (US National Research Council, 2011). The fact that many asteroids (i.e. the parent bodies of chondritic meteorites) are undifferentiated, and therefore largely unaltered since the solar system formed, is especially important in this respect. Moreover, the differentiated asteroids (i.e. the sources of metallic and achondritic meteorites) are also of great scientific interest, as their study will shed light on our understanding of the earliest stages of planetary differentiation into cores, mantles and crusts. These scientific objectives can only be fully addressed in the context of an ambitious programme of space exploration able to conduct *in situ* investigations of, and sample collection and return to Earth from, a large number of different types of asteroid.

(2) Mitigating the impact hazard: Some asteroids, especially those near-Earth asteroids (NEAs) on Earth-crossing orbits, represent an impact hazard for our planet. Based on data from NASA's Wide-Field Infrared Survey Explorer (WISE) spacecraft, more than 20,000 NEA's larger than 100m across are thought to exist, and there are probably over a million smaller objects (Mainzer et al, 2011). Although the probability of a civilisation-destroying impact is low (no more than one is expected every 100,000 years), as the Tunguska (1908) and Chelyabinsk (2013) impacts remind us, the Earth is hit by asteroidal (and/or cometary) debris capable of causing significant damage and loss of life much more frequently. In order to better mitigate this threat we need to increase our knowledge of the numbers and orbits of NEAs (especially for the smaller sizes where current catalogues are incomplete), and start developing techniques to disrupt or deflect any objects found on Earth collision trajectories. Achieving that latter capability requires both a spacefaring capability able to visit asteroids at will (and possibly at short notice) in order to place explosive or propulsive devices on their surfaces, *and* a detailed understanding of the nature of asteroidal surfaces so that we can predict how they will react to such interventions.

(3) Utilizing extraterrestrial resources: Asteroids represent a significant potential resource of raw materials, both in support of continued space exploration activities and for the wider global economy (e.g. Martin, 1985; Hartmann, 1986; Lewis et al., 1993; see also other chapters in this book). Many NEAs are relatively easy to reach in energy terms and have very low surface gravities, which would minimise the cost of transferring materials extracted from them to the vicinity of the Earth. Moreover, for many of these objects nature has already performed significant refining, or at least beneficiation, for us. For example, metallic asteroids (which constitute a few percent of the NEA population) consist of essentially pure nickel-iron alloy, and although Earth has significant reserves of both these elements they may still be very useful in the context of future space development. Perhaps of greater interest is the fact that metallic asteroids also contain about one hundred parts per million of gold and platinum group elements (PGEs), which are of sufficiently high value (for example as industrial catalysts) that they may be worth importing directly to Earth (e.g. Kargel, 1994). At today's prices for these elements ($20,000 to $50,000 per kilogram) it follows that a single small metallic asteroid about 200 metres across could be worth of the order of $100 billion dollars. Thus, in addition to being metaphorical scientific gold mines, some asteroids may prove to be *literal* gold mines as well! Moreover, although essentially rocky objects, ordinary chondritic asteroids (which probably account for the majority of NEAs) themselves consist of several percent Ni-Fe metal, which similarly contains hundred ppm-levels of PGEs. In addition, carbonaceous chondrites (which make up perhaps 10-15% of NEAs) are relatively rich in volatiles, which could be of great value to a future space economy by providing water, hydrogen, and oxygen for future space missions without the need to haul these materials out of Earth's gravity. Last but not least, there are also strong environmental arguments for mining even relatively common materials (such as iron, nickel, copper, and the increasingly important rare earth elements) from asteroids as an alternative to invasive strip-mining on Earth – asteroids do not have indigenous ecosystems that may be disrupted by mining activities whereas our planet does (see the discussion by Hartmann, 1986). For all these reasons, developing the capability of extracting useful resources from asteroids, and from other extraterrestrial sources, can be seen as an important investment in the future of the world economy (e.g. Crawford, 1995).

This book focusses mainly on the latter of these motivations, but clearly there are strong synergies between all three. For example, before either impact mitigation or economic utilization can be implemented it will be necessary to learn a great deal more about the nature and compositions of NEAs, both in general and for specific objects of interest. This implies initiating a major programme of scientific investigation, using both astronomical remote-sensing and *in situ* spacecraft investigations.

Moreover, the potential synergies also act the other way. As Martin Elvis (2012, see also his chapter in this book) has convincingly argued, the human and robotic missions needed to fully explore the Solar System, as well as the next generation of space telescopes required to advance our knowledge of the wider universe, will be so expensive that they may be unaffordable unless additional sources of funding can be identified. Leveraging some of the economic wealth locked up in NEAs (for example in the PGEs) may be one way to help finance ambitious future space exploration activities. And, of course, it goes without saying that the kind of infrastructural investments that will be required to extract the economic wealth of asteroids will be essentially the same as those required to destroy or deflect Earth-impacting asteroids should this ever prove to be necessary.

All these activities would benefit from greater international cooperation in space exploration by the World's space agencies, and the recognition that asteroids are important targets for human and robotic exploration. In this respect it is heartening that in September 2011 twelve of the world's space agencies came together, under the auspices of the International Space Exploration Coordination Group (ISECG), to produce a Global Exploration Roadmap for the human and robotic exploration of the inner Solar System (ISECG, 2011). In addition to missions to the Moon and Mars, which are important exploration targets in their own right, the Roadmap includes a strong focus on NEAs. Implementation of the Global Exploration Roadmap would both contribute to, and benefit from, the economic utililization of asteroidal resources.

Much also depends on the extent to which the cost of access to space can be reduced by new launch service providers such as SpaceX, or by new generations of vehicles such as the Skylon re-useable space-plane concept (e.g. Hempsell, 2010). The more the cost of accessing space can be reduced the greater will be the incentives for entrepreneurs to invest, and the sooner the economic utilization of asteroidal resources is likely to begin. By providing raw materials for *in situ* resource utilization (ISRU) for future space activities, as well as possible additional financing arising from the sale of valuable raw materials, asteroid mining may greatly facilitate ambitious space exploration activities such as envisaged by the Global Exploration Roadmap. Ultimately, it is not too much to hope that these activities will result in the creation of an economic and industrial infrastructure in the inner Solar System from which all humanity will benefit.

For all of these reasons, a book reviewing opportunities for the exploration and utilisation of asteroids is especially timely. Professor Viorel Badescu is therefore to be congratulated for compiling this volume, which I am sure will make a lasting contribution to the field.

Ian A. Crawford
Department of Earth and Planetary Sciences
Birkbeck College, University of London